\documentclass[preprint,12pt]{iopart}
\usepackage {iopams}  
\usepackage {mathrsfs}
\usepackage {graphicx}
\usepackage {dsfont}

\begin{document}
\paper[A fast search strategy for gravitational waves from LMXBs]
      {A fast search strategy for gravitational waves from low-mass X-ray binaries}

\author{C. Messenger and G. Woan}
\address{Department of Physics and Astronomy, University of Glasgow, Glasgow G12 8QQ, UK}
\ead{c.messenger@astro.gla.ac.uk}

\begin{abstract}
  We present a new type of search strategy designed specifically to find continuously emitting gravitational wave sources in known binary systems based on the incoherent sum
  of frequency modulated binary signal sidebands.  The search pipeline can be divided into three stages: the first is a wide bandwidth, $\mathcal{F}$-statistic search 
  demodulated for sky position.  This is followed by a fast second stage in which areas in frequency space are identified as signal candidates through the frequency
  domain convolution of the $\mathcal{F}$-statistic with an approximate signal template.  For this second stage only precise information on the orbit period and 
  approximate information on the orbital semi-major axis are required apriori.  For the final stage we propose a fully coherent Markov chain monte carlo based follow 
  up search on the frequency subspace defined by the candidates identified by the second stage.  This search is particularly suited to the low-mass X-ray binaries, for 
  which orbital period and sky position are typically well known and additional orbital parameters and neutron star spin frequency are not.  We note that for the 
  accreting X-ray millisecond pulsars, for which spin frequency and orbital parameters are well known, the second stage can be omitted and the fully coherent search stage 
  can be performed.  We describe the search pipeline with respect to its application to a simplified phase model and derive the corresponding sensitivity of the search. 
\end{abstract}


%
\section{Introduction}
%
The LIGO and GEO600 ground based interferometric gravitational wave detectors have been taking data in ``Science runs'' of ever-increasing length and sensitivity. 
since August 2002.  At present the detectors are midway through S5, a year long run started in November 2006, with LIGO 
running at design sensitivity and GEO600 not far behind.  Despite the unprecedented sensitivity of these instruments the data analysis task of identifying weak 
continuous gravitational wave (GW) signals buried in the detector noise represents a vast computational challenge.  To date a number of analyses have been performed 
using both LIGO and GEO600 data searching for targeted sources such as the known isolated and binary radio pulsars \cite{2004PhRvD..69h2004A,2005PhRvL..94r1103A}, and 
for wide parameter space searches including all-sky searches for isolated rapidly rotating neutron stars \cite{2005PhRvD..72j2004A,collaboration-2006} and searches for 
GWs from the low-mass X-ray binary (LMXB) Sco X-1 \cite{collaboration-2006,ballmer-2006-23}.  Present matched filter based search techniques, although optimal in terms 
of sensitivity for a given integration time, are unfeasibly computationally intensive due to the steep rise in the number of search templates with observation time.  As 
such, new less computationally intensive sub-optimal incoherent and hierarchical techniques must be developed whereby sacrifices in signal-to-noise ratio (SNR) can be 
recovered through longer observation times.  Examples of such methods include the stack-slide \cite{1998PhRvD..57.2101B}, Hough \cite{2005PhRvD..72j2004A}, and radiometer 
\cite{ballmer-2006-23} searches.  

We present a method previously employed in the electromagnetic detection of radio pulsars \cite{2003ApJ...589..911R} which we have adapted specifically for the detection 
of continuous GW sources in binary systems such as LMXBs.  This method, which we call the ``sideband'' search, is based on the incoherent summation of signal power 
present in the finite number of frequency modulated (FM) sidebands expected from such sources.  This ``sideband'' search stage is used to identify signal candidate 
frequencies forming a reduced frequency subspace on which a coherent Markov chain monte carlo (MCMC) based follow up can be run.  We identify that this follow up stage 
alone is suitable for searching for the accreting millisecond X-ray pulsars (AMXPs) 
\cite{1998Natur.394..344W,2002ApJ...575L..21M,2002ApJ...576L.137G,2003IAUC.8080....2M,2003IAUC.8144....1M,2005ApJ...622L..45G,2006ApJ...638..963K} for which the frequency 
is already known.  
 
We should stress that for the purposes of these proceedings we provide a description of this method applicable to the very basic case of circular orbit monochromatic 
sources of exactly known orbital period and sky position and assume a constant and flat detector noise spectrum within the band.  Here we aim to establish the principles 
behind this search strategy but for more general cases details of the practical application of this pipeline can by found in \cite{amp_paper,lmxb_paper}.  
%
\section{GW emission from LMXBs}
%
Observations by the Rossi X-ray Timing Explorer (RXTE) have provided evidence supporting the idea the that accretion torque supplied to neutron stars in accreting binaries 
could provide a sustained source of angular momentum able to power GW emission.  By balancing the accretion torque with the angular momentum radiated from the system 
through GWs we can estimate the GW amplitude received at Earth as  
\begin{equation}\label{eq:hbalance}
  h \sim 8.5\times 10^{-27}\left(\frac{r}{10\,\mathrm{kpc}}\right)^{-1}\left(\frac{f}{1000\,\mathrm{Hz}}\right)^{-\frac{1}{2}}\left(\frac{\dot{M}}{10^{-8}\,M_{\odot}\mathrm{yr}^{-1}}\right)^{\frac{1}{2}},
\end{equation}
where $r$ is the distance to the binary, $f$ is the gravitational wave frequency (twice the rotation frequency) and $\dot{M}$ is the accretion rate inferred from observed 
X-ray luminosity.  The RXTE observations showed an apparent clustering of spin frequencies in the range $250-370$ Hz well below the theoretical neutron star break-up 
frequency of $\sim 1$ kHz.  This implies the necessity for a mechanism through which an equilibrium can be achieved with the accretion torque.  The competing (non GW based) 
explanation for this possible equilibrium state is provided by an accretion disk - magnetosphere interaction model, \cite{1997ApJ...490L..87W} in which there is required 
a relation between accretion rate and neutron star magnetic field strength.  As this relation is not expected, the argument for GW emission from these systems remained
plausible \cite{1998ApJ...501L..89B} and has motivated the detailed investigation of the ability of a neutron star to support the non-zero quadrupole moment (or current 
quadrapole in the case of ``r-modes'' \cite{1998ApJ...502..708A}) needed for GW emission \cite{2000MNRAS.319..902U}.  More recently work by \cite{2005MNRAS.361.1153A} using 
a refined accretion model has been able to remove the need for an additional spin-down torque and although this therefore removes the need for GWs from these sources to 
explain the observed spin rates it does not make the GW generation mechanisms any less viable.

LMXBs are a class of semi-detached binary system consisting of a neutron star or black hole in orbit around a lower mass Roche lobe filling companion object, usually a white 
dwarf or brown dwarf star.  They have long been thought to be strong candidates for continuous gravitational wave emission due to the large ($\sim 100\%$ Eddington limit in 
some cases) accretion rates inferred from X-ray luminosity.  There are $85$ know LMXBs, \cite{2006yCat....102018R}, with periods ranging from $\sim 700$ sec to $\sim 33$ 
days and of these, $10$ have been seen to exhibit type 1 X-ray bursts \cite{2000AIPC..522..359B} allowing us to measure the spin frequency of these sources to an accuracy 
of $\sim 1$ Hz.  It has long been thought that the separation frequency between pairs of the kHz quasi-periodic oscillations (QPOs) seen in $17$ of the known LMXBs are 
directly related to the neutron star spin frequency \cite{2000ARA&A..38..717V} and recent observations of kHz QPOs in the millisecond accreting X-Ray pulsars 
SAX J1808.4-3658 \cite{2003Natur.424...44W} and XTE J1807-294 \cite{2005ApJ...634.1250L} have lent weight to this theory.  Unfortunately in LMXBs the kHz QPO separation 
frequencies are not constant and are seen to vary with source brightness.  Indeed for the $7$ LMXBs for which bursts and kHz QPO separation frequencies are known the simple 
relations obeyed by the AMXPs is not so clear cut. As a consequence we should assume that the unknown LMXB spin frequencies obey the same distribution described by those 
systems for which the frequency is known, currently ranging $250-620$ Hz.  This implies that any sensible and exhaustive search for these objects should use the entire 
ground-based interferometer sensitivity window of $\sim 100 - 1500$ Hz.
%
\section{The signal model}
%
We begin by assuming that the data received at a GW detector can be represented as the sum of the signal and additive Gaussian noise such that
\begin{equation}
  s(t) = W(t)\left[h(t) + n(t)\right],
\end{equation}
where the function $W(t)$ is a time domain window function required to describe the non-continuous operation of a gravitational wave detector and 
can be equal only to zero or unity at any given time.  We consider GWs emitted via quadrupole radiation from a rotating non-axisymmetric triaxial 
ellipsoid (GWs are emitted at twice the rotation frequency).  Following \cite{1998PhRvD..58f3001J} we write the signal as
\begin{equation}
  h(t) = \sum_{i=1}^{4}A_{i}h_{i}(t),
\end{equation}
where the signal amplitudes $A_{i}$ are
\begin{eqnarray}
  A_{1}&=&\left[h_{0}\left(1+\cos^{2}\iota\right)\cos{2\psi}\cos{\Phi_{0}}-2h_{0}\cos{\iota}\sin{2\psi}\sin{\Phi_{0}}\right]/2, \nonumber \\
  A_{2}&=&\left[h_{0}\left(1+\cos^{2}\iota\right)\sin{2\psi}\cos{\Phi_{0}}+2h_{0}\cos{\iota}\cos{2\psi}\sin{\Phi_{0}}\right]/2, \nonumber \\
  A_{3}&=&\left[-h_{0}\left(1+\cos^{2}\iota\right)\cos{2\psi}\sin{\Phi_{0}}-2h_{0}\cos{\iota}\sin{2\psi}\cos{\Phi_{0}}\right]/2, \nonumber \\
  A_{4}&=&\left[-h_{0}\left(1+\cos^{2}\iota\right)\sin{2\psi}\sin{\Phi_{0}}+2h_{0}\cos{\iota}\cos{2\psi}\cos{\Phi_{0}}\right]/2. \label{eq:A1234}
\end{eqnarray}
Here $h_{0}$ is the signal amplitude, $\iota$ is the inclination angle (the angle between the NS spin axis and the line of sight vector 
from source to detector), $\psi$ is the GW polarisation angle, and $\Phi_{0}$ is the signal phase at $t=t_{0}$.  The 4 
corresponding time dependent signal components $h(t)_{i}$ are
\begin{eqnarray}
  h_{1}=a(t)\cos{\Phi(t)}, &\hspace{1cm}& h_{3}=a(t)\sin{\Phi(t)},\nonumber \\
  h_{2}=b(t)\cos{\Phi(t)}, &\hspace{1cm}& h_{4}=b(t)\sin{\Phi(t)}.
\end{eqnarray}
where $\Phi(t)$ is the time dependent GW phase and the functions $a(t)$ and $b(t)$ are related to the GW detector antenna pattern functions
$F_{+}$ and $F_{\times}$ by 
\begin{equation}
  F_{+} = a(t)\cos\psi + b(t)\sin\psi,\hspace{0.5cm}F_{\times} = b(t)\cos\psi - a(t)\sin\psi.
\end{equation}
%
\subsection{A simple binary phase model}
%
We have limited our phase model to that of a monochromatic source in a circular binary orbit, so the corresponding GW phase is
\begin{equation}
  \Phi(t) =  2\pi f_{0}(t-t_{0}) + 2\pi f_{0}a\cos\left(2\pi(t-t_{p})/P\right),\label{eq:phi}
\end{equation}
where $f_{0}$ is the intrinsic (constant) GW frequency of emission, $a$ is the light crossing time of the orbital radius projected along the line of sight, $P$ is the 
orbital period, and $t_{p}$ represents a reference time at which the source passes through the ascending node of the orbit\footnote{For eccentric orbits $t_{p}$ represents 
the time of orbital periapsis, hence the subscript $p$.}.  We consider known systems for which the sky position is known to high accuracy and as such we assume that any phase 
contribution from the detector's motion with respect to the solar system barycentre (SSB) can be removed.  
%
\subsection{The Jacobi-Anger identity}
%
The chosen phase model in equation~\ref{eq:phi} shows that we are dealing with an FM signal, with modulation amplitude $2\pi f_{0}a$, and 
period equal to the orbital period $P$.  Using the Jacobi-Anger identity
\begin{equation}
  e^{\rmi z\cos\theta}=\sum_{n=-\infty}^{\infty}\rmi^{n}J_{n}(z)\rme^{\rmi n\theta},
\end{equation}
we can express this as an approximate complex phase factor 
\begin{equation}
  \rme^{\rmi\Phi(t)}\approx\sum_{n=-m}^{m}J_{n}(Z)\exp\left\{2\pi \rmi t\left(f_{0} + \frac{n}{P}\right)+\rmi n\left(\frac{\pi}{2}-\frac{t_{p}}{P}\right)\right\}, \label{eq:expphi1}
\end{equation}
where $J_{n}$ is the Bessel function of the first kind with argument $Z=2\pi f_{0}a$.  Note that we have reduced the problem from an infinite to a finite 
summation by taking advantage of the fact that $J_{n}(Z)\approx 0$ for $n>Z$.  In practise our approximation will be set to $m=Z=2\pi f_{0}a$.
%
\subsection{The $\mathcal{F}$-statistic}
\begin{figure}
  \begin{center}
    \includegraphics[width=10cm]{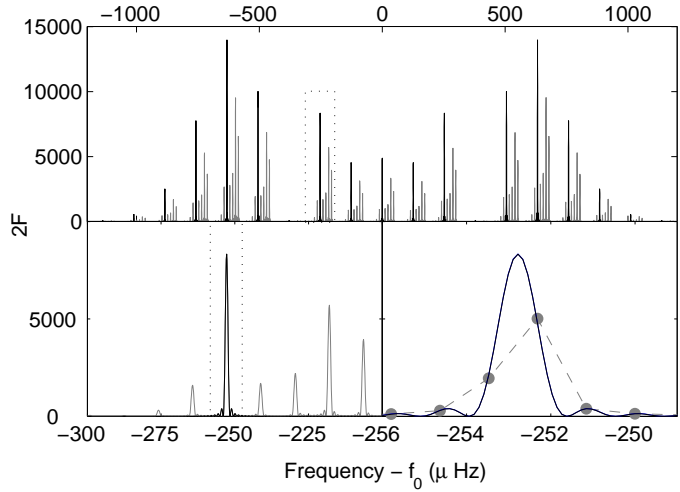}
    \caption{The $\mathcal{F}$-statistic as a function of frequency for a simulated signal with parameters $f_{0}=200 + 1.0849\times 10^{-6}$ Hz, 
      $a=0.005$ sec, $P=7912.7$ sec, $\alpha=2.0$ rad, $\delta=1.0$ rad, and nuisance parameters $h_{0}=1$, $\psi=0.2$ rad, $\cos\iota = 0.5$, 
      $\Phi_{0}=1$ rad.  The observation length is $T_{s}=T_{o}=10$ days with no gaps and a noiseless dataset (we have set $S_{h}=1$).  In the upper panel 
      we show $2\mathcal{F}$ (including AM sidebands) in solid grey, and $2\mathcal{F}$ neglecting the AM sidebands (as defined
      by equation~\ref{eq:twoF}) in solid black.  The dotted box in the upper panel indicates the area magnified shown in the lower left panel where we can 
      identify the amplitude modulation sidebands for this single frequency modulated sideband.  The dotted box in the lower left panel indicates the
      magnified area shown in the lower right panel where we have included a third evaluation of $2\mathcal{F}$ through direct computation of 
      equation~\ref{eq:Fstat} with frequency resolution $1/T_{s}$.}\label{fig:Fstat}
  \end{center}
\end{figure}

%
The $\mathcal{F}$-statistic \cite{1998PhRvD..58f3001J} is obtained through maximisation of the log likelihood with respect to the signal amplitudes $A_{i}$.  It is 
constructed from the complex components
\begin{eqnarray}
  F_{a}(f) &=& \int_{0}^{T_{s}}a(t)h(t)W(t)e^{-2\pi \rmi ft}\rmd t, \label{eq:Fa} \\
  F_{b}(f) &=& \int_{0}^{T_{s}}b(t)h(t)W(t)e^{-2\pi \rmi ft}\rmd t, \label{eq:Fb}
\end{eqnarray}
where the total observation time span is $T_{s}$ and the total observation time $T_{o}$ is $\int_{0}^{T_{s}}W(t)dt$.  The 
$\mathcal{F}$-statistic\footnote{Here we choose to use $2\mathcal{F}$ rather than $\mathcal{F}$ which is statistically more easily described.} itself is then 
defined as
\begin{equation}
  \fl 2\mathcal{F}(f) = \frac{8}{T_{o}S_{h}(f)D}\left[B|F_{a}(f)|^{2}+A|F_{b}(f)|^{2}-2C\Re\left(F_{a}(f)F_{b}(f)^{*}\right)\right] \label{eq:Fstat},
\end{equation}
where the quantities $A,B,C,D$ are
\begin{eqnarray}
  A = \frac{2}{T_{o}}\int_{0}^{T_{s}}a^{2}(t)W(t)\rmd t,\label{eq:A} &\hspace{1cm}& C = \frac{2}{T_{o}}\int_{0}^{T_{s}}a(t)b(t)W(t)\rmd t, \\
  B = \frac{2}{T_{o}}\int_{0}^{T_{s}}b^{2}(t)W(t)\rmd t,\label{eq:B} &\hspace{1cm}& D = AB-C^{2}, \label{eq:ABCD}
\end{eqnarray}
and $S_{h}(f)$ is the single-sided noise spectral density. By applying our phase model (equation~\ref{eq:phi}) we can write the components of the $\mathcal{F}$-statistic as
\begin{eqnarray}
  F_{a}(f)&=&\left(A+C\right)(A_{2}-\rmi A_{4})\sum_{n=-m}^{m}\frac{J_{n}(Z)}{2}\rme^{\rmi n\left(\frac{\pi}{2}-\frac{t_{p}}{P}\right)}\tilde{W}\left(f-f_{n}\right), \label{eq:FaQ} \\
  F_{b}(f)&=&\left(C+B\right)(A_{1}-\rmi A_{3})\sum_{n=-m}^{m}\frac{J_{n}(Z)}{2}\rme^{\rmi n\left(\frac{\pi}{2}-\frac{t_{p}}{P}\right)}\tilde{W}\left(f-f_{n}\right), \label{eq:FbQ}
\end{eqnarray}
where $f_{n}=f_{0}+n/P$ and we have neglected negative frequency components.  We have used the approximations
\begin{eqnarray}
  \int_{0}^{T_{o}}W(t)a^{2}(t)\rme^{-2\pi \rmi ft}\rmd t\approx \frac{1}{2}AT_{o}\tilde{W}(f), \label{eq:wat} \\
  \int_{0}^{T_{o}}W(t)b^{2}(t)\rme^{-2\pi \rmi ft}\rmd t\approx \frac{1}{2}BT_{o}\tilde{W}(f), \label{eq:wbt} \\
  \int_{0}^{T_{o}}W(t)a(t)b(t)\rme^{-2\pi \rmi ft}\rmd t\approx \frac{1}{2}CT_{o}\tilde{W}(f), \label{eq:wct} 
\end{eqnarray}
which are true for $T_{s}\gg P$ and at frequencies at or close to $f_{n}$.  We can now construct the $\mathcal{F}$-statistic via substitution 
of equations~\ref{eq:FaQ} and \ref{eq:FbQ} into equation~\ref{eq:Fstat} to obtain
\begin{equation}
  \fl 2\mathcal{F}\approx\frac{T_{o}}{2S_{h}}\left[A(A_{1}^{2}+A_{3}^{2})+B(A_{2}^{2}+A_{4}^{2})+2C(A_{1}A_{2}+A_{3}A_{4})\right]\sum_{n=-m}^{m}J_{n}^{2}(Z)|\tilde{W}(f)|^{2}. \label{eq:twoF}
\end{equation}
We have also chosen to neglect the amplitude modulation (AM) sidebands associated with each FM sideband for reasons of simplicity.  As can be seen 
from the example $\mathcal{F}$-statistic spectrum shown in figure~\ref{fig:Fstat}, there is significant additional $\mathcal{F}$-statistic located at the $8$ 
AM sidebands per FM sideband, each separated by $1/(\mathrm{sidereal\, day})$ Hz, and practical applications of this search technique should take advantage of 
this.    
%
\section{Searching for sources with unknown frequency - the sideband search}
%
The search for continuously emitting sources is generally made more difficult when the intrinsic spin frequency of the source is unknown,
and this is the case for LMXBs.  We shall now describe a method for searching for these sources based on the incoherent summation of sideband 
$\mathcal{F}$-statistic.  This technique is a fast way to search broad frequency bands for sources with known sky position and orbital 
period.  
%
\subsection{The sideband detection statistic}
%
The majority of $\mathcal{F}$-statistic values from a source in a binary system are spread broadly across a relatively wide bandwidth 
($\sim 4\pi f_{0}aT_{s}/P$ frequency bins) but are actually only present significantly in $\sim 4\pi f_{0}a$ isolated regions, each separated by $1/P$ Hz.  
We will define
\begin{equation}\label{eq:template}
  q(f) = \sum_{n=-m}^{m}\delta(f - f'_{n})
\end{equation}  
as an approximate search template where $f'_{n}$ is the frequency of the closest frequency bin to the frequency of the $n$th sideband, equal to $f_{n}=f_{0}+n/P$.
We have therefore constructed a comb of unit amplitude ``spikes'' each separated by $\approx1/P$ Hz.  We then convolve this template with the output of a 
wideband (sky position demodulated) $\mathcal{F}$-statistic search, $2\mathcal{F}(f)$\footnote{In practise this is most efficiently evaluated via the convolution 
theorem.} giving us
\begin{equation}\label{eq:Cf}
  C(f) = 2\mathcal{F}(f)\otimes q(f).
\end{equation}
We should note that the number of ``spikes'' within the template is a function of the orbital radius and that an optimal value of $C(f)$ will be obtained if using
its true value.  However, the power of this detection statistic is sensitive only to large $\sim 50\%$ mismatches between template orbital radius and the
true value and is therefore not required as an exactly known parameter.  This procedure is many orders of magnitude faster than the equivalent coherent matched filtered 
search but, as we shall describe in the following section, this increase in speed comes at the cost of reduced sensitivity.  An example of the sideband statistic 
$C(f)$ is shown in figure~\ref{fig:posteriors}.
%
\subsection{The sideband search sensitivity}
\begin{figure}
  \begin{center}
    \includegraphics[width=8cm]{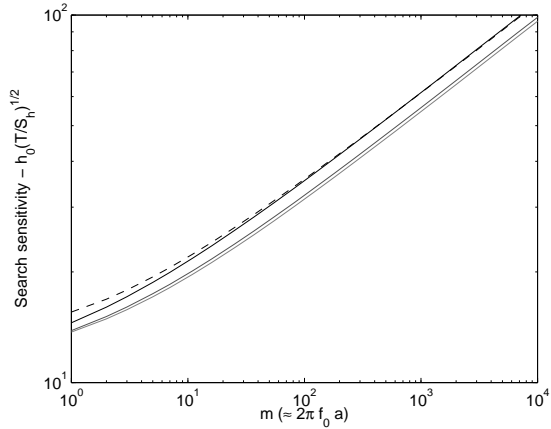}
    \caption{The sideband search sensitivity parameterised by $h_{0}\sqrt{T/S_{h}}$ for a $1\%$ false alarm rate and $10\%$ false dismissal rate averaged over the sky 
      and as a function of $m$.  We show in solid black, dark grey, and light grey, the curves corresponding to frequency resolutions of $1/T$, $1/2T$ and $1/4T$ 
      respectively.  The dashed black curve represents our approximation to the solid black curve using Gaussian distributions to model noise and signal plus noise 
      distributions.  Note that for large $m$ we see that the parameterised sensitivity is $\propto m^{1/4}$.}\label{fig:sensitivity}
  \end{center}
\end{figure}

%
The detection statistic $C(f)$ (equation~\ref{eq:Cf}) is simply the incoherent summation of sideband $\mathcal{F}$-statistics.  In an idealised case of data 
free from gaps and with constant noise amplitude, our window function $W(f)$ can be expressed analytically and the detection statistic becomes 
\begin{equation}\label{eq:Cfideal}
  C(f)=\frac{1}{2}\sum_{n=-m}^{m}\left\{\frac{1-\cos\left[2\pi(\Delta f_{n})T\right]}{\left[\pi(\Delta f_{n})\right]^{2}}\right\}2\mathcal{F}(f_{n}).
\end{equation}
We have used $\Delta f_{n}=f_{n}-f'_{n}$ to represent the sideband frequency mismatches due to our finite frequency resolution.  In practise 
$\Delta f_{0}$ will be drawn from a uniform distribution between the limits $[-\delta f/2 - \delta f/2)$, and for the purposes of this analysis we shall assume that 
this is the case for $\Delta f_{n}$.  For the noise only case, $2\mathcal{F}_{n}$ is a $\chi^{2}_{4}$-distributed random variable with mean and variance equal to $4$ 
and $8$ respectively.  The simple summation of $M=2m+1$ such independent random variables results in a $\chi^{2}_{4M}$-distributed
random variable.  The mean and variance of $C$ for noise alone are then
\begin{eqnarray}
  \mu_{0} (C) &=& 4M, \label{eq:munoise}\\
  \sigma^2_{0} (C) &=& 8M, \label{eq:sigmanoise}
\end{eqnarray} 
respectively.  When a signal is present in the noise, $2\mathcal{F}_{n}$ becomes a  non-centrally $\chi^{2}_{4}$-distributed with a non-centrality parameter 
$\lambda_{n}$ as a function of the random variable $\Delta f_{n}$.  In order to approximate the distribution of $C$ in this case we use
\begin{equation}\label{eq:sigapprox}
  2\mathcal{F}_{n}\left[\lambda_{n}(\Delta f_{n})\right]\approx |\tilde{W}(\Delta f_{n})|^{2}2\mathcal{F}_{n}(\lambda_{n}),
\end{equation}
where we have assumed that for values of $|\tilde{W}(\Delta f_{n})|^{2}$ close to unity the change in the product of the $\mathcal{F}$ statistic with the window function
is proportional to the relative change in the distribution mean.  From this approximation it follows that the mean and variance for signal plus noise are 
\begin{eqnarray}
  \fl \mu_{1}(C) &\approx& \rho_{\mathrm{opt}}^{2}\langle |\tilde{W}|^{2} \rangle + 4M, \label{eq:musignal}\\
  \fl \sigma^2_{1} (C) &\approx& \left(\frac{\rho_{\mathrm{opt}}^{4}}{M}+12\rho_{\mathrm{opt}}^{2} + 24M\right)\left(\frac{\rho_{\mathrm{opt}}^{4}\langle |\tilde{W}|^{4}\rangle + 8\rho_{\mathrm{opt}}^{2}\langle |\tilde{W}|^{2} \rangle +16M^{2}}{\rho_{\mathrm{opt}}^{4} + 8\rho_{\mathrm{opt}}^{2} +16M^{2}}\right) \nonumber \\ 
  \fl &&- 8\rho_{\mathrm{opt}}^{2}\langle |\tilde{W}|^{2} \rangle - 16M,\label{eq:sigmasignal}
\end{eqnarray} 
respectively, where
\begin{equation}\label{eq:optimalsnr}
  \sum_{n=-m}^{m}\lambda_{n} = \rho^{2}_{\mathrm{opt}} = \frac{2}{S_{h}(f_{0})}\int_{0}^{T_{s}}h^{2}(t)W(t)\rmd t.
\end{equation}
and $\langle\ldots\rangle$ indicates an averaging over the range of possible values of $\Delta f_{n}$.  By the central limit theorem we use the mean and variance 
approximations, in the case 
of signal plus noise, to describe the distributions of $C$ as Gaussian.  Figure.~\ref{fig:sensitivity} shows the relationship between search sensitivity, parameterised as 
$h_{0}\sqrt{T/S_{h}}$, as a function of $m$ and for various choices of frequency resolution $\delta f$.  
We can also use Equations~\ref{eq:munoise},~\ref{eq:sigmanoise},~\ref{eq:musignal} and \ref{eq:sigmasignal} to provide a rough guide to the sensitivity by approximating the SNR achievable with this 
search as
\begin{equation}\label{eq:SNRapprox}
  \rho\sim \frac{\mu_{1}(C)-\mu_{0}(C)}{\sqrt{\sigma^{2}_{1}(C)}}\sim \rho_{\mathrm{opt}}^{2}\sqrt{\frac{M}{\rho_{\mathrm{opt}}^{4}+12M\rho_{\mathrm{opt}}^{2} + 8M^{2}}}.
\end{equation}
For $M\gg\rho_{\mathrm{opt}}^{2}$ we see for a given sideband search SNR that
\begin{equation} \label{eq:h0prop}
  h_{0}\propto M^{1/4}\sqrt{S_{h}/T}.
\end{equation}
We should note that although this is an incoherent search technique, we still attain an $h_{0}$ sensitivity scaling proportional to $T^{-1/2}$, unlike most 
other incoherent searches.  This is due to the fact that $M\approx 2(2\pi f_{0}a)+1$ is a constant for a given source and therefore the $M^{1/4}$ factor in 
equation~\ref{eq:h0prop} remains constant for any choice of observation time.  The clear comparison here is to the stack-slide search \cite{1998PhRvD..57.2101B}
where the statistical behaviour of the detection statistic is very similar to the behaviour of the sideband statistic.  For stack-slide however, the analogue 
of the parameter $M$ is the number of stacks, which have typically predefined length making the number of stacks proportional to the total observation time 
and consequently a $T^{-1/4}$ scaling is achieved.   
%
\section{A Follow up exploration of the $\mathcal{F}$-statistic likelihood}
%
We propose using an MCMC approach with our $\mathcal{F}$-statistic signal model components (equations~\ref{eq:FaQ} and \ref{eq:FbQ}) to explore the associated 
likelihood space.  This is a natural extension to the sideband search in which the $\mathcal{F}$-statistic and its components have already been computed from our 
GW dataset.  By selecting values of the previous ``sideband'' search output, $C(f)$, above a predefined threshold, we identify isolated subspaces in our broadest 
parameter space dimension.  These subspaces correspond to signal candidate frequencies and it is over this subset of frequencies that we perform the MCMC.
%
\subsection{An MCMC exploration of the $\mathcal{F}$-statistic likelihood}
%
In order to perform an MCMC we must calculate the likelihood.  Let us first re-parameterise each $n$th complex $\mathcal{F}$-statistic components $F_{a}^{(n)}$ and 
$F_{b}^{(n)}$ as 
\begin{equation}\label{eq:Fredef}
  F_{a}^{(n)} = \mathds{F}_{1}^{(n)} + i\mathds{F}_{3}^{(n)}, \hspace{1cm} F_{b}^{(n)} = \mathds{F}_{2}^{(n)} + i\mathds{F}_{4}^{(n)},
\end{equation}
where $n$ indexes the FM sidebands $(-m\leq n \leq m)$.  For data containing signals in the presence of Gaussian noise the real variables $\mathds{F}^{(n)}_{i}$ are 
Gaussian distributed with means given by
\begin{eqnarray}
  \fl \bar{\mathds{F}}_{1}^{(n)} = J_{n}(Z)T(AA_{1}+CA_{2})/4, \hspace{0.5cm} \bar{\mathds{F}}_{3}^{(n)} = -J_{n}(Z)T(AA_{3}+CA_{4})/4, \nonumber \\
  \fl \bar{\mathds{F}}_{2}^{(n)} = J_{n}(Z)T(CA_{1}+BA_{2})/4, \hspace{0.5cm} \bar{\mathds{F}}_{4}^{(n)} = -J_{n}(Z)T(CA_{3}+BA_{4})/4, \label{eq:newFmeans}
\end{eqnarray}
and governed by a covariance matrix
\begin{equation}
  \mathrm{cov}(\mathds{F}_{i}^{(n)},\mathds{F}_{j}^{(n)}) = \left(\begin{array}{cc}
    \mathcal{C} & \mathcal{O} \\
    \mathcal{O} & \mathcal{C} \\
    \end{array}\right), \hspace{1cm}
  \mathcal{C} = \frac{TS_{h}}{8}\left(\begin{array}{cc}
    A & C \\
    C & B \\
    \end{array}\right) \label{eq:covariance},
\end{equation}
where $\mathcal{O}$ indicates a null $2\times 2$ matrix.  Therefore, the log likelihood function of the vector of $\mathcal{F}$-statistic components becomes
\begin{eqnarray}
  \fl\mathcal{L}(\vec{\mathds{F}}) &\propto& -\sum_{n=-m}^{m}B\left[(\mathds{F}_{1}^{(n)}-\bar{\mathds{F}}_{1}^{(n)})^{2} +  (\mathds{F}_{3}^{(n)}-\bar{\mathds{F}}_{3}^{(n)})^{2}\right] \nonumber \\ 
  \fl &&+A\left[(\mathds{F}_{2}^{(n)}-\bar{\mathds{F}}_{2}^{(n)})^{2} + A(\mathds{F}_{4}^{(n)}-\bar{\mathds{F}}_{4}^{(n)})^{2}\right] \nonumber \\
  \fl &&-2C\left[(\mathds{F}_{1}^{(n)}-\bar{\mathds{F}}_{1}^{(n)})(\mathds{F}_{2}^{(n)}-\bar{\mathds{F}}_{2}^{(n)}) + (\mathds{F}_{3}^{(n)}-\bar{\mathds{F}}_{3}^{(n)})(\mathds{F}_{4}^{(n)}-\bar{\mathds{F}}_{4}^{(n)})\right].\label{eq:loglikelihood}
\end{eqnarray}
We use an MCMC to efficiently explore this likelihood surface and to generate marginal posterior probability density functions (PDFs) for each of our search parameters.
By using values of $C(f)$ above a given threshold we can think of this as an approximate identification of small volumes of parameter space $V_{i}$ likely to be 
associated with large values of the likelihood (or log-likelihood).  By now treating $f_{0}$ as a search parameter within the MCMC we use $V_{i}$ as an approximation to 
our prior knowledge on that space by setting $p(f_{0}\notin V_{i})=0$.  All subspaces $V_{i}$ are then considered to form a continuous space on frequency such that a 
positive frequency jump from within subspace $V_{i}$ large enough to fall just outside its upper boundary will land in subspace $V_{i+1}$.  This way a single MCMC can 
be run and yield easily interpretable PDFs on our search space as opposed to separate PDFs for each frequency candidate which would leave us unable to make global parameter 
space inferences.

\begin{figure}
  \begin{center}
    \includegraphics[width=10cm]{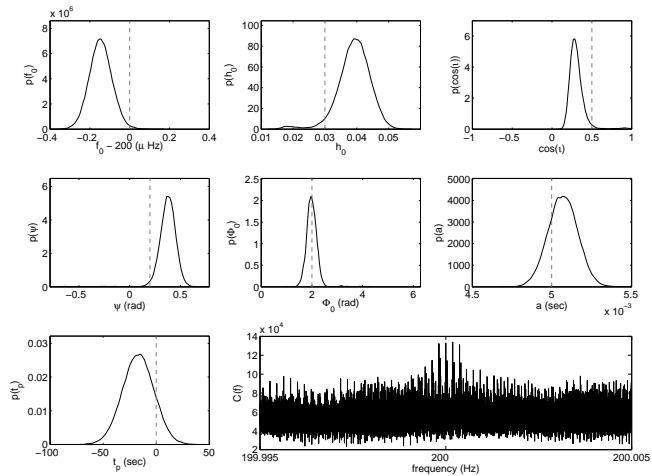}
    \caption{The marginal posterior PDFs (solid black lines) for an example LMXB type problem and the corresponding sideband detection statistic (lower right).  The broken 
      grey lines in the PDF plots indicate the true parameter values.  The detector noise is assumed known at $S_{h}=1$ and the observation time is $10$ days which for the 
      chosen signal parameters results in an optimal SNR of $?$.  The search band is $0.01$ Hz from which the $20$ largest values of $C$ were used as candidates for the
      MCMC prior frequency ranges.}\label{fig:posteriors}
  \end{center}
\end{figure}

For our simple LMXB model we define the primary search parameters as $\Lambda=\left\{f_{0},h_{0},\cos\iota,\psi,\phi_{0},a,t_{p}\right\}$.  
The form of the signal as described in equations~\ref{eq:FaQ} and \ref{eq:FbQ} shows us that the majority of $\mathcal{F}$-statistic is localised in $M$ sidebands
at frequencies $f_{0}+n/P$.  In the simplest cases this allows us to compute the log likelihood using only $4M$ data points (ie. each of the $4$ $\mathcal{F}$-statistic
components sampled at the closest discrete frequency bin $f'_{n}$ to each sideband).  Data used between sidebands would add only an identical constant factor to each 
computation of the log-likelihood for all $\Lambda$ thus giving no useful discriminating information.  By not using inter-sideband data the 
burden of the computation of $\mathcal{L}$ is therefore reduced by a factor of $\sim T/P$ and is therefore independent of observation time. 
We should also note that for our simplistic phase model each computation of $\mathcal{L}$ requires only a recalculation of the $\mathcal{F}$-statistic component means 
(equation~\ref{eq:newFmeans}) for each of the $M$ sidebands.  Figure.~\ref{fig:posteriors} shows a set of marginal posterior distributions for an example LMXB type problem.
We should also note that this MCMC stage alone is suited to the search for GWs from AMXPs for which the frequency and orbital parameters are known apriori to relatively 
high accuracy. 
%
\section{Summary}
%
We have presented a new incoherent search technique designed for performing GW searches for known objects in binary systems with unknown or poorly known frequencies,
specifically LMXBs.  Accurate knowledge of the orbital period allows the incoherent summation of frequency modulated $\mathcal{F}$-statistic sidebands removing orbital 
phase as a search parameter and also relaxing the requirement for accurate knowledge of orbital radius.  Of course, all of these benefits come at the price of reduced 
sensitivity to GW amplitude which we have shown to scale as $h_{0}\propto (4\pi f_{0}a)^{1/4}T^{1/2}$ implying that greatest sensitivity is achieved for short period 
(small orbital radius) binaries.  Full implementation of the search pipeline will take advantage of the multi-interferometer $\mathcal{F}$-statistic 
\cite{2005PhRvD..72f3006C} whereby prior to the summation of sideband power data sets from $N$ independent detectors are combined 
coherently thereby increasing sensitivity by $\sim \sqrt{N}$ for little additional computational cost.  We have also identified that the coherent follow up MCMC stage 
constitutes a complete search if applied to the AMXPs, a similar class of source to the LMXBs but for which the frequency and orbital parameters are well constrained 
apriori. We again stress that here we have applied the proposed pipeline only to a simple source model and neglected issues such as poorly known orbital period and sky 
position, non-zero eccentricity, realistic detector noise characteristics and more complex issues such as the detailed spin frequency evolution of accreting systems.  
However, we believe that solutions to these problems can be integrated into this search pipeline and are described in detail in \cite{amp_paper,lmxb_paper}.  Finally,
we stress that although the final search stage is a coherent one, the ultimate sensitivity of the complete pipeline is limited by the threshold set in the incoherent 
``sideband'' search stage.

\ack 

We wish to thank the members of the LSC pulsar working group for helpful discussions and their contributions to the LAL and LALapps data analysis packages. \\

\bibliographystyle{unsrt}
\bibliography{masterbib}

\end{document}